\documentstyle[twoside,fleqn,espcrc2,amsfonts]{article}

\newcommand{\AmS}{{\protect\the\textfont2
  A\kern-.1667em\lower.5ex\hbox{M}\kern-.125emS}}

\newcommand{\beq}{\begin{equation}}
\newcommand{\eeq}{\end{equation}}

\title{Ground State Entropy in Potts Antiferromagnets and Chromatic
       Polynomials} 

\author{R. Shrock$^{a*}$ and S.-H. Tsai\address{
        Institute for Theoretical Physics \\
        State University of New York, Stony Brook, NY 11794, USA}%
 \thanks{email: shrock@insti.physics.sunysb.edu.
        This research was supported in part by the NSF grant PHY-97-22101.
Present address for S.-H. Tsai: Dept. of Physics and Astronomy, Univ. of
        Georgia, Athens, GA  30602.}}
       
\begin{document}

\begin{abstract}

We discuss recent results on ground state entropy in Potts antiferromagnets and
connections with chromatic polynomials.  These include rigorous lower and upper
bounds, Monte Carlo measurements, large--$q$ series, exact solutions, and
studies of analytic properties.  Some related results on Fisher zeros of Potts
models are also mentioned.

\end{abstract}

\maketitle

\section{Introduction}

    Nonzero ground state entropy, $S_0 \ne 0$, is an important subject in
statistical mechanics, as an exception to the third law of thermodynamics 
(e.g. \cite{al}).  A physical example is ice.  A simple model exhibiting
ground state entropy is the $q$-state Potts antiferromagnet (PAF)
on a lattice, or more generally a graph $G$, for sufficiently
large $q$.  The zero-temperature partition function satisfies
\beq
Z(G,q,T=0)_{PAF}=P(G,q)
\label{zp}
\eeq where $P(G,q)$ is the chromatic polynomial expressing the number of ways
of coloring the vertices of $G$ with $q$ colors such that no two adjacent
vertices have the same color.  Let $\{G\} = \lim_{n \to \infty} G$.  The ground
state degeneracy per site $W$, given by $S_0 = k_B \ln W$, is 
\beq 
W(\{G\},q) =
\lim_{n \to \infty} P(G,q)^{1/n}
\label{w}
\eeq 
There are very few nontrivial exact solutions for $W$. 
We have obtained a number of new results on $W$, including rigorous bounds, 
large--$q$ series expansions, Monte Carlo measurements, exact
solutions, and studies of analytic properties \cite{p3afhc}-\cite{hs}. 
Besides physical (positive integral) $q$, it is of interest to consider $W$ as
a function of complex $q$; $W$ is an analytic function of $q$ except on a
continuous locus ${\cal B}$ (${\cal B}$ may be null, and there may also be 
isolated singularities of $W$).  As $n
\to \infty$, the locus ${\cal B}$ forms via coalescence of a subset of zeros of
$P(G,q)$.  $W$ is determined via (\ref{w}) in the region $R_1$ reached by
analytic continuation from the real $q$ axis for $q > \chi(G)$, where $\chi(G)$
is the chromatic number of $G$, i.e., the minimum number of colors needed to 
color $G$ with the above constraint. In other regions separated from $R_1$ 
by nonanalytic
boundaries comprising ${\cal B}$, only $|W|$ can be determined.  There is a
subtlety in the definition of $W$, since at certain
special points one encounters the noncommutativity of limits \cite{w} 
\beq
\lim_{n \to \infty} \lim_{q \to q_s} P(G,q)^{1/n} \ne
\lim_{q \to q_s} \lim_{n \to \infty} P(G,q)^{1/n}
\label{wnoncomm}
\eeq
At such points, we use the second order of limits to define $W$.  Physically,
one finds $W > 1$, i.e., $S_0 > 0$ for $q > \chi(G)$. 

\section{Bounds, Series, and Monte Carlo Measurements}

We have proved rigorous lower and upper bounds on $W$ for a number of lattices
$\Lambda$ \cite{ww,wn}. As an example, for the honeycomb $(hc)$ lattice, we 
get 
\beq
W(hc,q)_\ell \le W(hc,q) \le W(hc,q)_u
\label{whcbound}
\eeq
where
\beq
W(hc,q)_\ell = \frac{(q^4-5q^3+10q^2-10q+5)^{1/2}}{q-1}
\label{whclower}
\eeq
and 
\beq
W(hc,q)_u = (q^2-3q+3)^{1/2}
\label{wchupper}
\eeq 
These bounds are very restrictive even for moderate $q$, as is clear from
the fact that the first three terms in a large--$q$ expansion coincide. 
Although a bound on a given function need not, {\it a priori},
coincide with a series expansion of that function, we find that the lower bound
(\ref{whclower}) coincides with the first eleven terms of the large--$q$
expansion for $W(hc,q)$.  

Accordingly, we have extended our study of the lower bound and have discovered
and proved a generalization applicable to the full set of Archimedean lattices
\cite{wn}.  An Archimedean lattice is a uniform tiling of the plane with one or
more regular polygons such that all vertices are equivalent to each other.  It
can be specified by the ordered sequence of polygons $p_i$ traversed by a 
circuit around any vertex: 
$\Lambda = \prod_i p_i^{a_i}$.  Let $\sum a_i = a_{i,s}$ and $\nu_i=
a_{i,s}/p_i$.  Then our general lower bound is 
\beq
W \Bigl ( (\prod_i p_i^{a_i}),q \Bigr ) \ge 
\frac{\prod_i D_{p_i}(q)^{\nu_{p_i}}}{q-1}
\label{wlbarch}
\eeq
with
\beq
D_k(q) = 
\sum_{s=0}^{k-2}(-1)^s {{k-1}\choose {s}} q^{k-2-s}
\label{dk}
\eeq
We have calculated large--$q$ series expansions for a number of Archimedean
lattices and have compared the lower bounds with these series \cite{wn}.  
For the square, triangular, and honeycomb lattices we have
carried out Monte Carlo measurements of $W(\Lambda,q)$ for $q$ values
up to 10 and have found that even for moderate values of $q$, the upper and
lower bounds bracket the measured values quite closely \cite{ww}. 

\section{Analytic Structure of $W(\{G\},q)$}

We have calculated exact solutions for $W(\{G\},q)$ for a number of families
of graphs and have studied their analytic structure.  
A general form for $P(G,q)$ is 
\beq
P(G,q) =  c_0(q) + \sum_{j=1}^{N_a} c_j(q)(a_j(q))^{t_j n}
\label{pgsum}
\eeq where $c_j(q)$ and $a_j(q)$ are certain functions of $q$. Here the
$a_j(q)$ and $c_{j \ne 0}(q)$ are independent of $n$, while $c_0(q)$ may
contain $n$-dependent terms, such as $(-1)^n$, but does not grow with $n$ like
$(const.)^n$ with $|const.| > 1$.  A term $a_\ell(q)$ is leading if it
dominates the $n \to \infty$ limit of $P(G,q)$; in particular, if $N_a \ge 2$,
then it satisfies $|a_\ell(q)| \ge 1$ and $|a_\ell(q)| > |a_j(q)|$ for $j \ne
\ell$, so that $|W|=|a_\ell|^{t_j}$.  The locus ${\cal B}$ occurs where there
is a nonanalytic change in $W$ as the leading terms $a_\ell$ in
eq. (\ref{pgsum}) changes.  For a given $\{G\}$ one can ask various questions
about ${\cal B}$: (i) does it separate the $q$ plane into separate regions?
(ii) is it compact or noncompact?  (iii) how many disjoint components does it
contain? (iv) does it have multiple points where several branches cross? (v)
does it cross the positive real axis or contain a segment lying along this
axis, and if so, what is the maximal real value, $q_c$, in ${\cal B}$?  We have
answered these questions for various families \cite{w,wc,wn,wa,hs}. 

For example, such families as circuit and ladder graphs yield ${\cal
B}$ that do separate the $q$ plane into various regions.  With M. Ro\v{c}ek, we
have used a generating functional method to obtain exact solutions for $W$ on
infinitely long, finite-width homogeneous strip graphs \cite{strip}.  For these
we find that the loci ${\cal B}$ consist of arcs that do not separate regions. 

In general, the zeros of $P(G,q)$ only merge to form ${\cal B}$ in the $n \to
\infty$ limit, but for an interesting set of families, they actually lie
exactly on ${\cal B}$ even for finite $n$.  We have constructed an infinite set
of these families, called $p$-wheels, $(Wh)^{(p)}_n = K_p + C_{n-p}$, where
$K_r$ is the ``complete'' graph (each of whose vertices is connected to all
others), $C_r$ is the circuit graph, and $G+H$ connotes joining of vertices of
$G$ to those of $H$.  We proved that the zeros of $P((Wh)^{(p)}_n,q)$ lie
precisely on ${\cal B}$, which is the circle $|q-(p+1)|=1$ \cite{wc}.  This is
reminiscent of the Yang-Lee circle theorem \cite{yl} although different in
certain respects.

For $\{G\}$ with noncompact ${\cal B}$ passing through $z=1/q = 0$, no
large--$q$ series expansions exist.  Since these series are very useful for
regular lattices, it is important to understand the properties of a family
$\{G\}$ that yield a noncompact ${\cal B}$.  We have given a general condition
for ${\cal B}$ to be noncompact and have constructed a number of these families
\cite{wa}; one simple infinite set of families is $G = (K_p)_b + H$,
where $b$ signifies the removal of $b$ bonds from a vertex of $K_p$. 
For example, for $1 \le b \le p-1$ and $H=T_r$ (the tree graph on
$r$ vertices), in the limit $r \to \infty$, ${\cal B}$ is the vertical line
with $Re(q) = p+\frac{1}{2}$, or equivalently, in the $z=1/q$ plane, the circle
$|z-z_c|=z_c$ with $z_c=2/(2p+1)$.  Additional families can be generated by
various types of homeomorphic expansions of the basic $G = (K_p)_b + H$ 
(homeomorphic expansion = insertion of degree-2 vertex on a bond). 

\section{Approach to 2D thermodynamic limit}

Using exact solutions of $W$ for infinitely long strip graphs of finite width 
$L_y$, we have explored the approach of
$W$ to its 2D thermodynamic limit \cite{w2d}.  
We showed that the approach of $W$ to its 2D
thermodynamic limit as $L_y$ increases is quite rapid; for moderate values of
$q$ and $L_y \simeq 4$, $W(\Lambda,(L_x=\infty) \times L_y,q)$ is within about
${\cal O}(10^{-3})$ of the 2D value $W(\Lambda,(L_x=\infty) \times
(L_y=\infty),q)$ for periodic transverse boundary conditions (b.c.). 
The approach of $W$ to the 2D thermodynamic limit was proved 
to be monotonic (non-monotonic) for free (periodic) transverse b.c.  
An application to compute central charges for
cases with critical ground states was noted. 

\section{Complex-temperature properties}

We have calculated Fisher zeros of the partition function for the Potts model
on the square lattice for several $q$ values and related the inferred CT phase
boundaries to locations of singularities in the thermodynamic functions 
obtained from analyses of low-temperature series expansions
\cite{pf}.  These studies have been extended to honeycomb, triangular,
kagom\'e, and diced lattices \cite{p,p2}.  Although in general CT singularities
have rather different properties than physical critical singularities in spin
models, such as violation of universality \cite{chisq}, 
we have found an exact mapping that
relates the free energy of the $q$-state
Potts antiferromagnet on a lattice $\Lambda$ for the full temperature
interval $0 \le T \le \infty$ and the free energy of the $q$-state Potts
model on the dual lattice for a semi-infinite CT interval, 
$-\infty \le a_d \le -(q-1)$, where $a_d=(a+q-1)/(a-1)$ and $a=e^K$.  Hence,
for this interval, CT singularities of the free energy are equivalent to
physical critical singularities.  Effects of
next-nearest-neighbor couplings were studied in Ref. 
\cite{1dnnn}.

\end{document}